\documentclass[preprint,12pt]{revtex4}
\usepackage{amssymb}
\usepackage{graphicx}
\usepackage{bm}
\usepackage{epsfig}
\usepackage{color}

\usepackage{color}
\usepackage{amstext}

\begin{document}

\title{Excitation energy transfer from dye molecules to doped graphene}
\author{R. S. Swathi and K. L. Sebastian}
\affiliation{Department of Inorganic and Physical Chemistry\\
Indian Institute of Science, Bangalore 560012, India }
\date{\today}

\begin{abstract}
Recently, we have reported theoretical studies (J. Chem. Phys. 129,
054703, 2008 and J. Chem. Phys. 130, 086101, 2009) on the rate of
energy transfer from an electronically excited molecule to graphene.
It was found that graphene is a very efficient quencher of the
electronically excited states and that the rate $\propto $
$(distance)^{-4}$.   The process was found to be effective up to
$30\;nm$ which is well beyond the traditional FRET limit. In this
report, we study the transfer of an amount of energy $\hbar \Omega$
from a dye molecule to doped graphene.  We find a crossover of the
distance dependence of the rate from $(distance)^{-4}$ to
exponential as the Fermi level is increasingly shifted into the
conduction band, with the crossover occurring at a shift of the
Fermi level by an amount $\hbar \Omega/2$.
\end{abstract}

\maketitle

\section{Introduction}

Excitation energy transfer involving carbon based materials is
interesting due to the fact that using such materials, it is
possible to measure distances well beyond the traditional FRET
limit. In our earlier papers, we have analyzed the process of
resonance energy transfer from an excited dye molecule to a sheet of
graphene \cite{jcp1, jcp2, jchemsci}. The rate was evaluated as a
function of the distance $z$ of the molecule from the graphene
sheet. We have found the process of energy transfer to be very
efficient and the rate has a $z^{-4}$ dependence on the distance.
Our report was the first study on energy transfer to graphene.
Recent experiments that have been performed after our theoretical
studies have infact found efficient energy transfer to graphene and
the process was found to be useful in identifying graphene flakes
both on substrates and in solution \cite{jacsfqm}. Quenching by
graphene was also found to be useful in obtaining good resonance
Raman signals from fluorescent samples \cite{jacsrrs}, in
fabricating graphene based devices \cite{nanotech} and in
quantitative DNA analysis \cite{advfunctmater, analchem}. We have
also studied the process of energy transfer from fluorophores to
carbon nanotubes and found a $d^{-5}$ dependence \cite{jcp3}.
Quantum chemical studies on energy transfer involving two carbon
nanotubes have also been reported \cite{scholes}. All the above
studies involve energy transfer to extended charge densities of
carbon based materials and hence found a deviation from the
$(distance)^{-6}$ dependence, which has been obtained within the
dipolar approximation. Such deviations from the dipolar
approximation have also been found in polymers \cite{rossky},
quantum wells \cite{klimov} etc. The Fermi surface of undoped
graphene is a set of six points known as the K-points. As a result
of this, the density of states at the Fermi level is zero. It is
possible to shift the Fermi level of graphene away from the K-point
experimentally, either by electrical or chemical doping
\cite{natmat, natphys}. This will make the density of states at the
new Fermi level non-zero. In this letter we study the effect of
shifting the Fermi level on the distance dependence of the rate of
energy transfer to graphene. We imagine that the Fermi level is
shifted into the conduction band to a level with magnitude of wave
vector, $k_F$. To keep the calculations simple, we use the Dirac
cone approximation, which allows us to get analytical expressions
for the rate at large distances.  We note that as we are shifting
the Fermi level by rather large amounts, there will be sizeable
corrections to the rate due to deviations from the Dirac cone
approximation and hence our conclusions are of qualitative nature.

\section{Model for the rate}

\begin{figure}[tbp]\centering
\includegraphics[width=0.7\linewidth,height=0.6\linewidth]{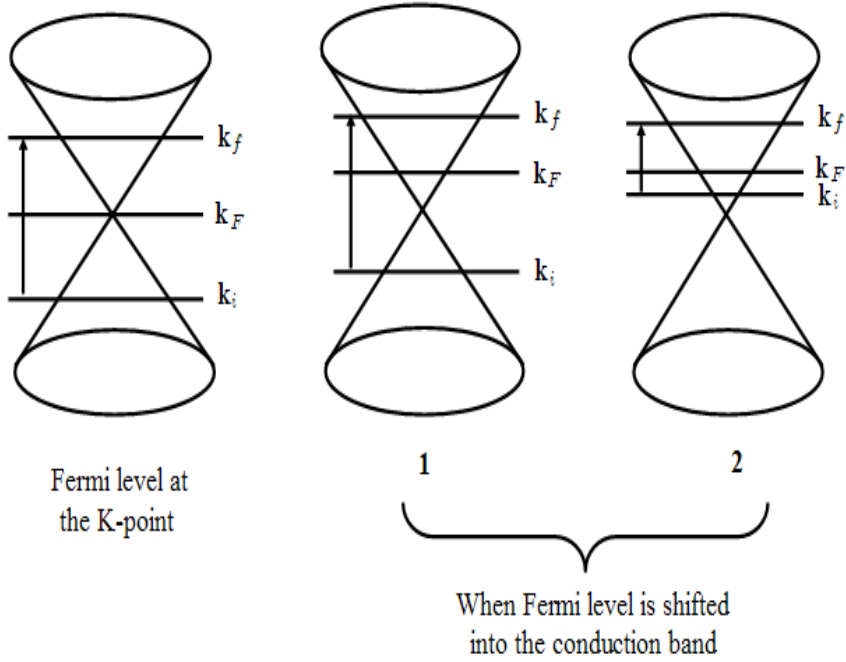}\newline
\caption{A schematic of the energy bands, showing the Fermi wave
vectors, the initial and the final wave vectors corresponding to
excitation energy transfer. $E_F=v_f k_F$ is the location of the new
Fermi level on doping graphene. } \label{doping}
\end{figure}

We consider the process of excitation energy transfer from a dye
molecule to doped graphene. Since the energy donor (dye molecule)
has a localized electronic charge density, we think of the
interaction between the donor and the acceptor as that between the
transition
dipole of the donor, $\bm{\mu }_{eg}^{D}$, given by $\bm{\mu }_{eg}^{D}=-e\int d\mathbf{%
r}_{1}\psi _{e}^{D\ast }\left( \mathbf{r}_{1}\right)
\mathbf{r}_{1}\psi _{g}^{D}\left( \mathbf{r}_{1}\right) $ and the
transition charge density $\rho \left( \mathbf{r}_2\right) =-e\psi
_{g}^{A\ast }\left( \mathbf{r}_{2}\right) \psi _{e}^{A}\left( \mathbf{r}%
_{2}\right)$ of the acceptor \cite{jchemsci, jcp3}. The matrix
element for interaction is given by
\begin{equation}
U=\bm{\mu }_{eg}^{D}\cdot \nabla \Phi ,  \label{intenergy}
\end{equation}%
where $\Phi $ is the electrostatic potential at the point
$\mathbf{r}$ (the position of the donor) due to the charge density
$\rho \left( \mathbf{r}_2\right)$ and is given by
\begin{equation}
\Phi \left( \mathbf{r}\right) =\frac 1{4\pi \epsilon}\int
d\mathbf{r}_2\frac{\rho \left( \mathbf{r}_2\right) }{\left|
\mathbf{r}-\mathbf{r}_2\right| } \label{pphi}.
\end{equation}
As a result of energy transfer, an electron in graphene with wave
vector $\mathbf{k}_i$ is excited to a level with
wave vector $\mathbf{k}_f$. We define $\mathbf{k}_{f}=\mathbf{k}_{i}+\mathbf{q}$, where $%
\mathbf{q}\hbar $ is the momentum transferred to graphene. The rate
of energy transfer can be evaluated using the Fermi golden rule and
is given by
\begin{equation}
k=\frac{2\pi }{\hbar }\sum\limits_{\mathbf{k}%
_{i}}\sum\limits_{\mathbf{q}}\mid U_{\mathbf{k}_{i},\mathbf{q}}\mid
^{2}\delta
(E_{\mathbf{k}_{i}+\mathbf{q}}^+-E_{\mathbf{k}_{i}}^--\hbar \Omega
). \label{ratekiq2}
\end{equation}
We use the tight binding wave functions of graphene and evaluate the
matrix element (for details, see \cite{jcp1,jcp2,jchemsci}) using
Eqs. (\ref{intenergy}) and (\ref{pphi}) and find it to be
\begin{equation}
U =\frac {e}{4 \epsilon
A}\left[ e^{i(\delta _{%
\mathbf{k}_{i}+\mathbf{q}}-\delta
_{\mathbf{k}_{i}})}-1\right] \bm{\mu}%
_{eg}^D\cdot \left( i\hat {\mathbf{q}}+\hat {\mathbf{k}}\right)
e^{-qz}e^{-i\mathbf{q}\cdot\mathbf{ X}} \label{uu},
\end{equation}
where $\hat {\mathbf{q}}=\frac{ \mathbf{q}}q$ is the unit vector in
the direction of $\mathbf{q}$ and $\hat {\mathbf{k}}$ is the unit
vector in the $z$ direction. We have also used $\mathbf{r}=\left(
\mathbf{X},z\right) $, with $\mathbf{X}$ being parallel to the plane
of graphene. $A$ is the area of the graphene lattice and
\begin{equation} \delta _{\mathbf{k}}=\tan ^{-1}\left(
\frac{k_{y}}{k_{x}}\right)=\varphi _{\mathbf{k}}, \label{dk}
\end{equation}%
where $\varphi _{\mathbf{k}}$ is the angle that the vector
$\mathbf{k}$ makes with the x-axis. We substitute Eq. (\ref{uu}) for
the interaction energy into Eq. (\ref{ratekiq2}) to get
\begin{equation}
k=\frac{\pi e^2}{4 \hbar \epsilon^2A^2}\sum_{\mathbf{q}}
|\mathbf{\mu}^D_{eg}.(i\mathbf{\hat \mathbf{q}}+\mathbf{\hat
\mathbf{k}})   |^2 \exp(-2qz)G(\mathbf{q}) \label{kk},
\end{equation}
where
\begin{equation}
G\left(\mathbf{ q}\right)=\sum\limits_{\mathbf{k_i}}\left[ 1-\cos \left( \varphi _{\mathbf{k}%
_{i}+\mathbf{q}}-\varphi _{\mathbf{k}_{i}}\right) \right]\delta
(E_{\mathbf{k}_i+\mathbf{ q}}^+-E_{\mathbf{k}_{i}}^--\hbar \Omega ).
\end{equation}
When the Fermi level of graphene is shifted into the conduction
band, the rate of energy transfer has contributions from two
different sets of transitions. In the first, $\mathbf{k}_i$ lies in
the valence band with $0\leq k_i\leq \infty $ and $\mathbf{k}_f$
lies in the conduction band with $k_F< k_f< \infty $. In the second,
both $\mathbf{k}_i$ and $\mathbf{k}_f$ lie in the conduction band
with $0\leq k_i\leq k_F$ and $k_F< k_f< \infty $ (see Fig.
\ref{doping}). The total rate can thus be written as a sum total of
both the contributions, $k=k_1+k_2$. $k_1$ and $k_2$ are both given
by
\begin{equation}
k_i=\frac{\pi e^2}{4 \hbar \epsilon^2A^2}\sum_{\mathbf{q}}
|\mathbf{\mu}^D_{eg}.(i\mathbf{\hat \mathbf{q}}+\mathbf{\hat
\mathbf{k}})   |^2 \exp(-2qz)G_i(\mathbf{q}) \label{kki},
\end{equation}
but with differing expressions for $G_i\left(\mathbf{ q}\right)$.
\begin{equation}
G_1\left(\mathbf{ q}\right)=\sum\limits_{\mathbf{k_i}\in\; valence\; band }\left[ 1-\cos \left( \varphi _{\mathbf{k}%
_{i}+\mathbf{q}}-\varphi _{\mathbf{k}_{i}}\right) \right]\delta
(E_{\mathbf{k}_i+\mathbf{ q}}^+-E_{\mathbf{k}_{i}}^--\hbar \Omega
)\Theta \left( \left| \mathbf{k}_i+\mathbf{q}\right| -k_F\right) .
\end{equation}
The theta function is introduced to satisfy the condition, $k_F<
k_f< \infty $. In a similar fashion,
\begin{equation}
G_2\left(\mathbf{ q}\right)=\sum\limits_{\mathbf{k_i}\in\; conduction\; band}\left[ 1+\cos \left( \varphi _{\mathbf{k}%
_{i}+\mathbf{q}}-\varphi _{\mathbf{k}_{i}}\right) \right]\delta
(E_{\mathbf{k}_i+\mathbf{ q}}^+-E_{\mathbf{k}_{i}}^+-\hbar \Omega
)\Theta \left( \left| \mathbf{k}_i+\mathbf{q}\right|
-k_F\right)\Theta \left(k_F-\left| \mathbf{k}_i\right|\right)
\label{g2q}.
\end{equation}
The two theta functions satisfy the conditions, $0\leq k_i\leq k_F$
and $k_F< k_f< \infty $. We now evaluate $G_1\left(\mathbf{
q}\right)$ and $G_2\left(\mathbf{ q}\right)$ separately. We replace
the sum over $\mathbf{k}_i$ in the expression for $G_1\left(\mathbf{
q}\right)$ by an integral and use the linear dispersion relation for
the energy levels of graphene ($E_{\mathbf{k}}^\pm=\pm v_f k$) to
get
\begin{equation}
G_1\left(\mathbf{ q}\right)=\frac{A}{4\pi ^2 v_f}\int
d\mathbf{k}_{i}\left[ 1-\frac{\mathbf{k}_i\cdot \left(
\mathbf{k}_i+\mathbf{q}\right) }{\left| \mathbf{k}_i\right| \left|
\left( \mathbf{k}_i+\mathbf{q}\right) \right| } \right]\delta
(\left| \mathbf{k}_{i}\right| + \left| \mathbf{k}_{i}+
\mathbf{q}\right|-\frac{\hbar \Omega}{v_f} )\Theta \left( \left|
\mathbf{k}_i+\mathbf{q}\right| -k_F\right).
\end{equation}
Introducing a new variable $\mathbf{ k}_{i}^{^{\prime }}$ defined by
$\mathbf{k}_{i}^{^{\prime }}=\mathbf{k}_{i}+\frac{ \mathbf{q}}{2}$
leads to
\begin{equation}
\begin{array}{c}
G_1\left(\mathbf{ q}\right)=\frac{A}{4\pi ^2 v_f}\int
d\mathbf{k}_{i}^{^{\prime }}\left[ 1-\frac{\left(
\mathbf{k}_{i}^{^{\prime }}-\frac{ \mathbf{q}}{2}\right)\cdot \left(
\mathbf{k}_{i}^{^{\prime }}+\frac{ \mathbf{q}}{2}\right) }{\left|
\mathbf{k}_{i}^{^{\prime }}-\frac{ \mathbf{q}}{2}\right| \left|
\mathbf{k}_{i}^{^{\prime }}+\frac{ \mathbf{q}}{2}\right| }
\right]\delta \left(\left| \mathbf{k}_{i}^{^{\prime }}-\frac{
\mathbf{q}}{2}\right| + \left| \mathbf{k}_{i}^{^{\prime }}+\frac{
\mathbf{q}}{2}\right|-\frac{\hbar \Omega}{v_f} \right) \times \\
\Theta \left( \left| \mathbf{k}_{i}^{^{\prime }}+\frac{
\mathbf{q}}{2}\right| -k_F\right).
\end{array}
\end{equation}
We choose the direction of $\mathbf{q}$ as the x-axis and then make
another change of variable to $\mathbf{r}$ given by
$\mathbf{r}=\frac{\mathbf{k}_{i}^{^{\prime }}}{q/2}$ to get
\begin{equation}
\begin{array}{c}
G_1\left(\mathbf{ q}\right)=\frac{A q^2}{16\pi ^2 v_f}\int
d\mathbf{r}\left[ 1-\frac{\left(
\mathbf{r}-\hat{\mathbf{i}}\right)\cdot \left(
\mathbf{r}+\hat{\mathbf{i}}\right)
}{\left|\mathbf{r}-\hat{\mathbf{i}}\right| \left|
\mathbf{r}+\hat{\mathbf{i}}\right| } \right]\delta \left[
\frac{q}{2} \left(\left| \mathbf{r}-\hat{\mathbf{i}}\right| + \left|
\mathbf{r}+\hat{\mathbf{i}}\right|\right)-\frac{\hbar \Omega}{v_f}
\right]\times \\  \Theta \left(\frac{q}{2}\left|
\mathbf{r}+\hat{\mathbf{i}}\right|  -k_F\right).
\end{array}
\end{equation}
Using $\mathbf{r}$ $ \equiv (x,y) $, the above equation can be
rewritten as
\begin{equation}
\begin{array}{c}
G_1\left(\mathbf{ q}\right)=\frac{A q^2}{8\pi ^2
v_f}\int\limits_{-\infty }^\infty dx\int\limits_0^\infty dy\left[
1-\frac{x^2+y^2-1}{\sqrt{\left( x-1\right) ^2+y^2}\sqrt{\left(
x+1\right) ^2+y^2}}\right] \times \\ \delta \left[ \frac q2\left(
\sqrt{\left( x-1\right) ^2+y^2}+\sqrt{\left( x+1\right)
^2+y^2}\right) -\frac{\hbar \Omega }{v_f}\right] \Theta
\left(\frac{q}{2} \sqrt{\left( x+1\right) ^2+y^2}-k_F\right).
\end{array}
\end{equation}
We now change over to elliptic coordinates defined by $x=\mu
\upsilon $ and $y=\sqrt{\left( \mu ^2-1\right) \left( 1-\upsilon
^2\right) }$. The transformation gives $dxdy=\frac{\mu ^2-\upsilon
^2}{\sqrt{\left( \mu ^2-1\right) \left( 1-\upsilon ^2\right) }}d\mu
d\upsilon $. With the above transformation, we get
\begin{equation}
G_1\left(\mathbf{ q}\right)=\frac{A q^2}{4\pi ^2
v_f}\int\limits_1^\infty d\mu \int\limits_{-1}^1d\upsilon
\sqrt{\frac{1-\upsilon ^2}{\mu ^2-1}}\delta \left( q\mu -\frac{\hbar
\Omega }{v_f}\right)\Theta \left[\frac{q}{2}\left(\mu+\upsilon
\right) -k_F\right].
\end{equation}
The integral over $\mu$ can be easily performed to get
\begin{equation}
G_1\left(\mathbf{ q}\right)=\frac{A q^2}{4\pi ^2 }\frac{\Theta
(\hbar \Omega -qv_{f})}{\sqrt{ \left( \hbar \Omega \right)
^{2}-q^{2}v_{f}^{2}}}\int\limits_{-1}^1d\upsilon \sqrt{1-\upsilon
^2} \Theta \left[\frac{q}{2}\left(\frac{\hbar \Omega}{q
v_f}+\upsilon \right) -k_F\right].
\end{equation}
The above equation can be rewritten as
\begin{equation}
G_1\left(\mathbf{ q}\right)=\frac{A q^2}{4\pi ^2 }\frac{\Theta
(\hbar \Omega -qv_{f})}{\sqrt{ \left( \hbar \Omega \right)
^{2}-q^{2}v_{f}^{2}}}\Theta \left[1 -\frac{2 E_F-\hbar
\Omega}{qv_f}\right] \int\limits_{Max[-1,\frac{2 E_F-\hbar
\Omega}{qv_f}]}^1d\upsilon \sqrt{1-\upsilon ^2} .
\end{equation}
We now substitute back the above expression into the rate expression
of Eq. (\ref{kki}) and convert the sum over $\mathbf{q}$ to an
integral to get
\begin{equation}
\begin{array}{c}
k_1=\frac{e^2}{64 \hbar \epsilon^2 \pi^3}\int\limits_{0}^{\infty}dq
q^3 e^{-2qz}\frac{\Theta (\hbar \Omega -qv_{f})}{\sqrt{ \left( \hbar
\Omega \right) ^{2}-q^{2}v_{f}^{2}}}\Theta \left[1 -\frac{2
E_F-\hbar \Omega}{qv_f}\right] \int\limits_{Max[-1,\frac{2
E_F-\hbar \Omega}{qv_f}]}^1d\upsilon \sqrt{1-\upsilon ^2} \times \\
\int\limits_0^{2\pi }d\theta \left| \bm{ \mu} _{eg}^D\cdot \left(
i\hat {\mathbf{q}}+\hat {\mathbf{k}}\right) \right| ^2,
\end{array}
\end{equation}
where $(q, \theta)$ are the polar coordinates of $\mathbf{q}$. After
performing the integral over $\theta$, we average over all possible
orientations of the donor transition dipole (see \cite{jcp2,
jchemsci}) to get
\begin{equation}
\begin{array}{c}
k_1=\frac{e^2 \mu_{eg}^2}{48 \hbar \epsilon^2
\pi^2}\int\limits_{0}^{\infty}dq q^3 e^{-2qz}\frac{\Theta (\hbar
\Omega -qv_{f})}{\sqrt{ \left( \hbar \Omega \right)
^{2}-q^{2}v_{f}^{2}}}\Theta \left[1 -\frac{2 E_F-\hbar
\Omega}{qv_f}\right]\int\limits_{Max[-1,\frac{2 E_F-\hbar
\Omega}{qv_f}]}^1d\upsilon \sqrt{1-\upsilon ^2} .
\end{array}
\end{equation}
Evaluation of the integral  over $\upsilon$ leads to
\begin{equation}
\begin{array}{c}
k_1=\frac{e^2 \mu_{eg}^2}{96 \hbar \epsilon^2
\pi^2}\int\limits_{0}^{\infty}dq q^3 e^{-2qz}\frac{\Theta (\hbar
\Omega -qv_{f})}{\sqrt{ \left( \hbar \Omega \right)
^{2}-q^{2}v_{f}^{2}}}\Theta \left[1 -\frac{2 E_F-\hbar
\Omega}{qv_f}\right]\left[ \frac \pi 2-\left( u\sqrt{1-u^2}+\sin
^{-1}u\right) \right],
\end{array}
\end{equation}where $u$ is defined by $u=Max[-1,\frac{2 E_F-\hbar \Omega}{qv_f}]$.

We now evaluate $G_2\left(\mathbf{ q}\right)$, defined by Eq.
(\ref{g2q}). Using the same procedure as before, we find that
\begin{equation}
G_2\left(\mathbf{ q}\right)=\frac{A q^2}{4\pi ^2
v_f}\int\limits_1^\infty d\mu \int\limits_{-1}^1d\upsilon
\sqrt{\frac{\mu ^2-1}{1-\upsilon ^2}}\delta \left( q\upsilon
-\frac{\hbar \Omega }{v_f}\right)\Theta
\left[\frac{q}{2}\left(\mu+\upsilon \right) -k_F\right]\Theta
\left[k_F-\frac{q}{2}\left(\mu-\upsilon \right) \right].
\end{equation}
The integral over $\upsilon$ can be easily performed to get
\begin{equation}
G_2\left(\mathbf{ q}\right)=\frac{A q^2}{4\pi ^2 }\frac{\Theta (
qv_{f}-\hbar \Omega)}{\sqrt{q^{2}v_{f}^{2}- \left( \hbar \Omega
\right) ^{2}}}\int\limits_{1}^\infty d\mu \sqrt{\mu ^2-1} \Theta
\left[\frac{q}{2}\left(\frac{\hbar \Omega}{q v_f}+\mu \right)
-k_F\right]\Theta \left[k_F-\frac{q}{2}\left(\mu-\frac{\hbar
\Omega}{q v_f} \right) \right].
\end{equation}
The above equation can be rewritten as
\begin{equation}
G_2\left(\mathbf{ q}\right)=\frac{A q^2}{4\pi ^2 }\frac{\Theta (
qv_{f}-\hbar \Omega)}{\sqrt{q^{2}v_{f}^{2}- \left( \hbar \Omega
\right) ^{2}}}\int\limits_{Max[1,\frac{2 E_F-\hbar
\Omega}{qv_f}]}^{\frac{2 E_F+\hbar \Omega}{qv_f}} d\mu \sqrt{\mu
^2-1} \Theta \left[\frac{2 E_F+\hbar \Omega}{qv_f}-1\right].
\end{equation}Substituting the above expression back into the rate expression of Eq.
(\ref{kki})  gives
\begin{equation}
\begin{array}{c}
k_2=\frac{e^2}{64 \hbar \epsilon^2 \pi^3}\int\limits_{0}^{\infty}dq
q^3 e^{-2qz}\frac{\Theta (qv_{f}-\hbar
\Omega)}{\sqrt{q^{2}v_{f}^{2}- \left( \hbar \Omega \right)
^{2}}}\Theta \left[\frac{2 E_F+\hbar \Omega}{qv_f}-1\right]
\int\limits_{Max[1,\frac{2 E_F-\hbar \Omega}{qv_f}]}^\frac{2
E_F+\hbar \Omega}{qv_f}d\mu \sqrt{\mu^2-1}
\times \\
 \int\limits_0^{2\pi }d\theta \left| \bm{ \mu} _{eg}^D\cdot
\left( i\hat {\mathbf{q}}+\hat {\mathbf{k}}\right) \right| ^2.
\end{array}
\end{equation}
The $\theta$ integral can be performed easily, followed by an
averaging over all possible orientations of the donor transition
dipole to get
\begin{equation}
\begin{array}{c}
k_2=\frac{e^2 \mu_{eg}^2}{48 \hbar \epsilon^2
\pi^2}\int\limits_{0}^{\infty}dq q^3 e^{-2qz}\frac{\Theta
(qv_{f}-\hbar \Omega)}{\sqrt{q^{2}v_{f}^{2}- \left( \hbar \Omega
\right) ^{2}}}\Theta \left[\frac{2 E_F+\hbar
\Omega}{qv_f}-1\right]\int\limits_{Max[1,\frac{2 E_F-\hbar
\Omega}{qv_f}]}^{\frac{2 E_F+\hbar \Omega}{qv_f}}d\mu \sqrt{\mu^2-1}
.
\end{array}
\end{equation}The integral over $\mu$ can be evaluated to get
\begin{equation}
\begin{array}{c}
k_2=\frac{e^2 \mu_{eg}^2}{96 \hbar \epsilon^2
\pi^2}\int\limits_{0}^{\infty}dq q^3 e^{-2qz}\frac{\Theta
(qv_{f}-\hbar \Omega)}{\sqrt{q^{2}v_{f}^{2}- \left( \hbar \Omega
\right) ^{2}}}\Theta \left[\frac{2 E_F+\hbar
\Omega}{qv_f}-1\right]\times \\
\left(-r\sqrt{r^2-1}+s\sqrt{s^2-1}+\log \left[ r+\sqrt{r^2-1}\right]
-\log \left[ s+\sqrt{s^2-1}\right] \right),
\end{array}
\end{equation}
where $r=Max[1,\frac{2 E_F-\hbar \Omega}{qv_f}]$ and $s=\frac{2
E_F+\hbar \Omega}{qv_f}$. The integrals over $q$ in the expressions
for $k_1$ and $k_2$ can be performed numerically, thus getting the
total rate of transfer when the Fermi level   in graphene is shifted
into the conduction band.

\section{Large $z$ behavior of $k_1$ and $k_2$}

It is easy to analyze the large $z$ behavior of $k_1$ and $k_2$ (see
the Appendix for the detailed analysis). When $E_F<\frac{\hbar
\Omega}{2}$ and $z>\frac{v_f}{2\Delta\epsilon}$, with $\Delta
\epsilon=\frac{\hbar \Omega}{2}-E_F$ and $\hbar \Omega>>\Delta
\epsilon$, the long-range behavior of $k_1$ is given by Eq.
(\ref{oo}) of the Appendix as
\begin{equation}
k_1=\frac{e^2 \mu_{eg}^2}{256 \pi \Omega  \hbar^2 \epsilon^2 z^4}.
\end{equation}
In the case when $E_F>\frac{\hbar \Omega}{2}$ and
$z>\frac{v_f}{4\Delta\epsilon}$, with $\hbar \Omega>>\Delta
\epsilon$,
\begin{equation}
k_1=\frac{e^2 \mu_{eg}^2  {\left| \Delta \epsilon \right|}^{3/2}}{96
\sqrt{2} \Omega \hbar^2 \epsilon^2 \pi^{3/2}v_f^{3/2} z^{5/2}}
e^{\frac{-4\left| \Delta \epsilon \right|z}{v_f}},
\end{equation}
as obtained in Eq. (\ref{ppp}) of the Appendix. The major
contribution from the $k_1$ term to the rate comes only when
$E_F<\frac{\hbar \Omega}{2}$ and it has a power law dependence
($z^{-4}$) on the distance. When $E_F>\frac{\hbar \Omega}{2}$, the
contribution from $k_1$ decreases exponentially with $z$ and hence
is very small. The long-range behavior of $k_2$ for both
$E_F<\frac{\hbar \Omega}{2}$ and $E_F>\frac{\hbar \Omega}{2}$ is
given by Eqs. (\ref{k2long}) and (\ref{k2long1}) of the Appendix as
\begin{equation}
\begin{array}{c}
k_2=\sqrt{\frac{e^4 \mu_{eg}^4 \hbar^3 \Omega^5}{ \pi^3 \epsilon^4
v_f^7}}\frac{e^{ \frac{-2 z\hbar\Omega}{v_f}}}{192\sqrt{z}}\left(
r\sqrt{r^2-1}-\log \left[ r+\sqrt{r^2-1}\right] \right),
\end{array}
\end{equation}
where $r=1+\frac{2 E_F}{\hbar \Omega}$.

Therefore, the large $z$ behavior of $k_2$ is exponential. Thus, in
the case when $E_F<\frac{\hbar \Omega}{2}$, the rate of transfer to
doped graphene has a power law dependence on the distance (arising
from the $k_1$ term), while when $E_F>\frac{\hbar \Omega}{2}$, the
rate has an exponential dependence (arising due to both $k_1$ and
$k_2$ terms). Therefore, as the Fermi level is increasingly shifted
into the conduction band, there is a crossover of the distance
dependence of the rate from power law to exponential and the
crossover occurs over a region of $E_F$ centred at $\frac{\hbar
\Omega}{2}$.

\begin{figure}[tbp]\centering
\includegraphics[width=0.7\linewidth,height=0.6\linewidth]{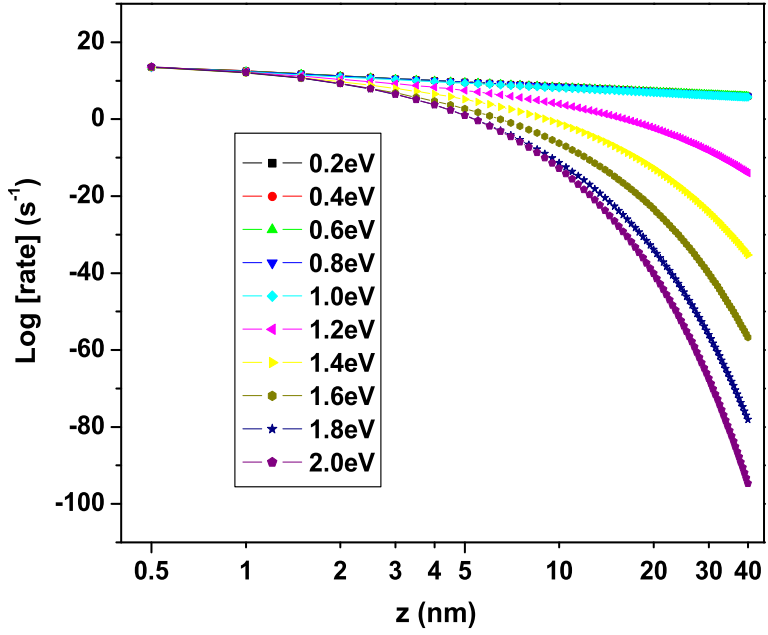}\newline
\caption{The rate of energy transfer as a function of distance, as
the Fermi level is shifted into the conduction band. The emission
energy of the fluorophore is taken to be $\hbar \Omega=2.0\;eV$.}
\label{data1}
\end{figure}

\begin{figure}[tbp]\centering
\includegraphics[width=0.7\linewidth,height=0.6\linewidth]{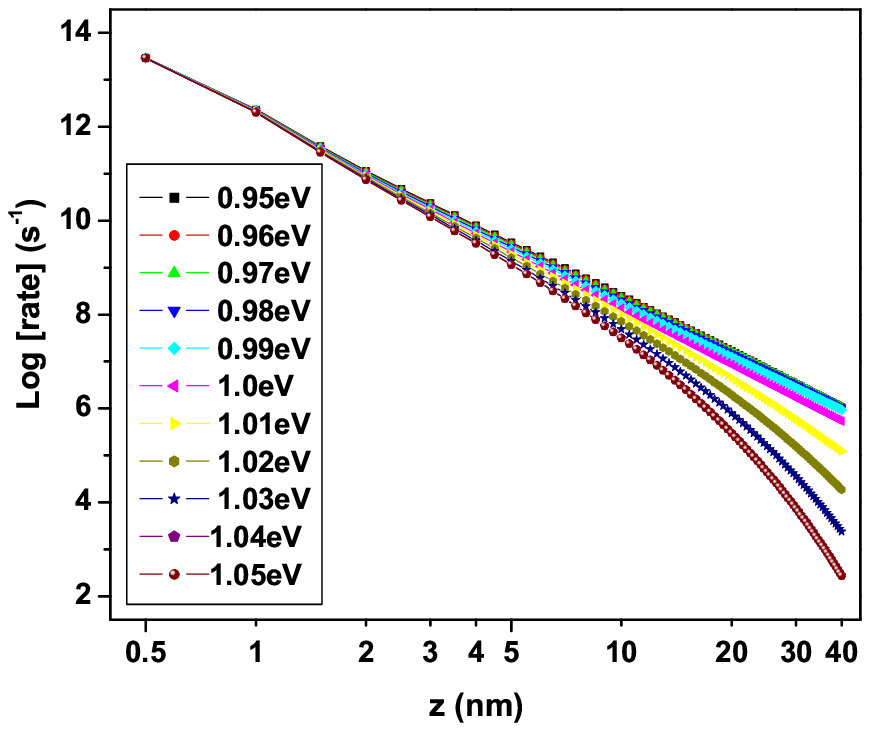}\newline
\caption{The rate of energy transfer as a function of distance, as
the Fermi level is shifted into the conduction band. The emission
energy of the fluorophore is taken to be $\hbar \Omega=2.0\;eV$. }
\label{data2}
\end{figure}

\begin{figure}[tbp]\centering
\includegraphics[width=0.7\linewidth,height=0.6\linewidth]{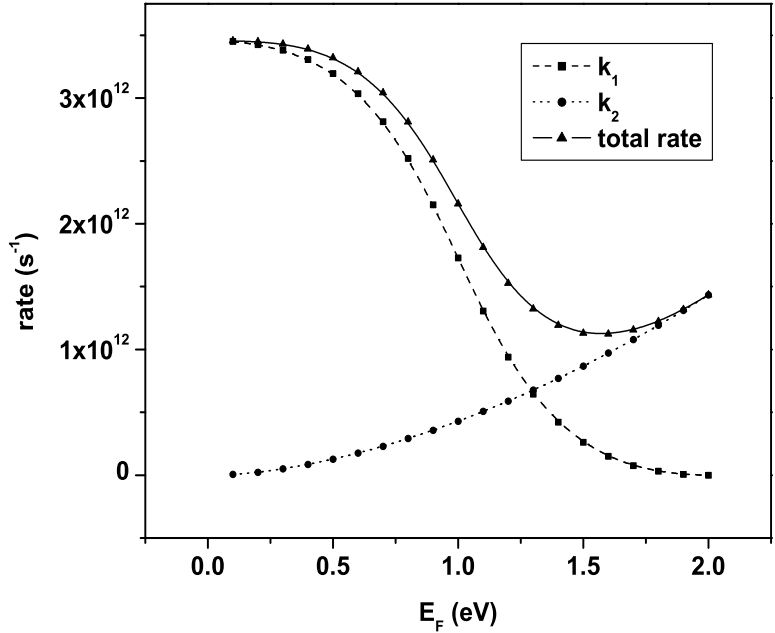}\newline
\caption{The rate of energy transfer as a function of the Fermi
energy of graphene, at a distance of $10\;\mathring{A}$. The figure
shows plots of $k_1$ and $k_2$ separately along with the total rate
of transfer.} \label{fermi}
\end{figure}

\section{Results}

We performed numerical calculations for evaluating the rates of
energy transfer from a fluorophore to doped graphene. We take the
emission energy of the fluorophore, $\hbar \Omega$ to be $2.0\;eV$.
Such low energy emission has been found in squarylium dyes
\cite{jacs}. We take $\mu_{eg}=4.5\;D$. Fig. \ref{data1} shows a
plot of the logarithm of the rate as a function of the logarithm of
the distance as the Fermi level is shifted increasingly into the
conduction band in the range $0.2-2.0\;eV$. The log-log plot is
linear showing that the rate has a power law dependence ($z^{-4}$)
on the distance for $E_F<1.0\;eV$. After around $1.0\;eV$, there is
deviation from linearity in the log-log plot and the dependence
becomes exponential. In order to look into the cross over region
more closely, in Fig. \ref{data2}, we show a plot of the rate as the
Fermi level is moved into the conduction band in the range
$0.95-1.05\;eV$. This clearly shows that there is a crossover of the
distance dependence of the rate from power law to exponential as the
Fermi level is increasingly shifted into the conduction band. It
should be possible to observe this effect experimentally.

Fig. \ref{fermi} shows a plot of the $k_1$ and $k_2$ terms, and the
total rate of transfer as a function of the Fermi energy of graphene
at a fixed distance, $z=10\;\mathring{A}$. The contribution from the
$k_1$ term decreases as the Fermi level is increasingly shifted into
the conduction band, while that from the $k_2$ term increases. From
the figure, it is clear that the total rate is governed mainly by
the $k_1$ term up to the crossover point. Beyond that, the
contribution from $k_1$ term is small and the total rate is governed
by the $k_2$ term. This is clearly a density of states effect.

\section{Conclusions}

We have studied the process of energy transfer from a fluorophore to
doped graphene. We have analyzed the distance dependence of the rate
of transfer as the Fermi level of graphene is shifted away from the
K-point into the conduction band. We find a crossover of the
dependence from power law ($z^{-4}$) to exponential as the Fermi
level is increasingly moved into the conduction band. The point of
crossover is at a shift of the Fermi level by $\hbar \Omega/2$.

\textbf{Acknowledgement}. R. S. Swathi acknowledges Council of
Scientific and Industrial Research (CSIR), India and
Bristol-Myers-Squibb fellowship for financial support. The work of
K.L. Sebastian was supported by the J.C. Bose Fellowship of the DST
(India).

%\bibliographystyle{apsrev}
%\bibliography{doping}

\section{Appendix}

\subsection{Behavior of rate constant, $k_1$}

  We first look at $k_{1}$. We consider two separate cases.
   \(\)   \text{\text{  }}\( \)   \begin{enumerate}
\item
Case I: We consider the case where $E_F$ is close to $\frac{\hbar
\Omega}{2}$, but less than it. We put $\Delta \epsilon=\frac{\hbar
\Omega}{2}-E_F$. $k_1$ is then given by
\begin{equation}
\begin{array}{c}
k_1=\frac{e^2 \mu_{eg}^2}{48 \hbar \epsilon^2
\pi^2}\int\limits_{0}^{\infty}dq q^3 e^{-2qz}\frac{\Theta (\hbar
\Omega -qv_{f})}{\sqrt{ \left( \hbar \Omega \right)
^{2}-q^{2}v_{f}^{2}}} \Theta \left[1 +\frac{2 \Delta
\epsilon}{qv_f}\right]\int\limits_{Max[-1,-\frac{2 \Delta \epsilon
}{qv_f}]}^1d\upsilon \sqrt{1-\upsilon ^2} \label{eqq}.
\end{array}
\end{equation}
As $\Delta \epsilon>0$, $\Theta \left[1 +\frac{2 \Delta
\epsilon}{qv_f}\right]=1$. For $z>\frac{v_f}{2\Delta \epsilon} $,
the major contribution to the above integral is from $q \in\
(0,\frac{\Delta \epsilon}{v_f})$. In this range,
\begin{equation}
\int\limits_{Max[-1,-\frac{2 \Delta \epsilon }{qv_f}]}^1d\upsilon
\sqrt{1-\upsilon ^2}=\int\limits_{-1}^1d\upsilon \sqrt{1-\upsilon
^2}=\frac{\pi}{2}.
\end{equation}
If $\frac{\hbar \Omega}{2}>>\Delta \epsilon$, then, in this range,
$(\hbar \Omega)^2-q^2 v_f^2\simeq (\hbar \Omega)^2$ and the integral
in Eq. (\ref{eqq}) may be approximated as
\begin{equation}
k_1=\frac{e^2 \mu_{eg}^2}{96 \pi \hbar^2 \epsilon^2
\Omega}\int\limits_{0}^{\infty}dq q^3 e^{-2qz}=\frac{e^2
\mu_{eg}^2}{256 \pi \Omega  \hbar^2 \epsilon^2 z^4}  \label{oo}.
\end{equation}

\item
Case II: We now consider the case $E_F>\frac{\hbar \Omega}{2}$.
Hence, $\frac{\hbar \Omega}{2}-E_F=-\left| \Delta \epsilon \right|$.
Then,
\begin{equation}
\begin{array}{c}
k_1=\frac{e^2 \mu_{eg}^2}{48 \hbar \epsilon^2
\pi^2}\int\limits_{0}^{\infty}dq q^3 e^{-2qz}\frac{\Theta (\hbar
\Omega -qv_{f})}{\sqrt{ \left( \hbar \Omega \right)
^{2}-q^{2}v_{f}^{2}}} \Theta \left[1 -\frac{2\left| \Delta \epsilon
\right| }{qv_f}\right]\int\limits_{Max[-1,\frac{2 \left| \Delta
\epsilon \right|  }{qv_f}]}^1d\upsilon \sqrt{1-\upsilon ^2} .
\end{array}
\end{equation}
For large $z$ $\left(z>>\frac{v_f}{4 \left| \Delta \epsilon
\right|}\right)$, this may be approximated as
\begin{equation}
k_1=\frac{e^2 \mu_{eg}^2}{48 \hbar \epsilon^2
\pi^2}\int\limits_{\frac{2 \left| \Delta \epsilon
\right|}{v_f}}^{\infty}dq q^3 e^{-2qz}\frac{\Theta (\hbar \Omega
-qv_{f})}{\sqrt{ \left( \hbar \Omega \right) ^{2}-q^{2}v_{f}^{2}}}
\int\limits_{\frac{2 \left| \Delta \epsilon \right|
}{qv_f}}^1d\upsilon \sqrt{1-\upsilon ^2} .
\end{equation}
For $z$ such that $\frac{4\left| \Delta \epsilon \right|z}{v_f}>>1$,
in the above integral over $q$, the contribution is from $q$ in the
vicinity of $\frac{2\left| \Delta \epsilon \right|}{v_f}$. We now
change the variable of integration from $q$ to $y$ defined by
$q=\frac{2\left| \Delta \epsilon \right|y}{v_f}$. Using the above
transformation, $k_1$ is given by
\begin{equation}
k_1=\frac{e^2 \mu_{eg}^2  {\left| \Delta \epsilon \right|}^4}{3
\hbar \epsilon^2 \pi^2v_f^4}\int\limits_{1}^{\infty}dy y^3
e^{\frac{-4\left| \Delta \epsilon \right|zy}{v_f}}\frac{\Theta
(\hbar \Omega -2\left| \Delta \epsilon \right|y)}{\sqrt{ \left(
\hbar \Omega \right) ^{2}-4\left| \Delta \epsilon \right|^2y^2}}
\int\limits_{\frac{1}{y}}^1d\upsilon \sqrt{1-\upsilon ^2} .
\end{equation}
Now, the major contribution to the above integral comes from values
of $y\simeq1$. If $y$ is close to unity,
\begin{equation}
\int\limits_{\frac{1}{y}}^1d\upsilon \sqrt{1-\upsilon
^2}\simeq\frac{2\sqrt{2}}{3}\left(\frac{y-1}{y}\right)^{3/2}.
\end{equation}
Hence,
\begin{equation}
k_1\simeq\frac{2\sqrt{2}e^2 \mu_{eg}^2  {\left| \Delta \epsilon
\right|}^4}{9 \hbar \epsilon^2 \pi^2v_f^4}\int\limits_{1}^{\infty}dy
y^{3/2}\left(y-1\right)^{3/2} e^{\frac{-4\left| \Delta \epsilon
\right|zy}{v_f}}\frac{\Theta (\hbar \Omega -2\left| \Delta \epsilon
\right|y)}{\sqrt{ \left( \hbar \Omega \right) ^{2}-4\left| \Delta
\epsilon \right|^2y^2}} .
\end{equation}
For $\frac{\hbar \Omega}{2}>>\left| \Delta \epsilon \right|$ and $y$
in the vicinity of $1$, $\Theta (\hbar \Omega -2\left| \Delta
\epsilon \right|y)=1$ and $\sqrt{ \left( \hbar \Omega \right)
^{2}-4\left| \Delta \epsilon \right|^2y^2}\simeq \hbar \Omega $. The
integral over $y$ can now be evaluated to get:
\begin{equation}
k_1\simeq\frac{e^2 \mu_{eg}^2  {\left| \Delta \epsilon
\right|}^2}{48 \sqrt{2} \Omega \hbar^2 \epsilon^2 \pi^2v_f^2 z^2}
e^{\frac{-2\left| \Delta \epsilon
\right|z}{v_f}}K_2\left(\frac{2\left| \Delta \epsilon
\right|z}{v_f}\right).
\end{equation}
For large values of $z$, the asymptotic form of the Bessel function
$K_2\left(\frac{2\left| \Delta \epsilon \right|z}{v_f}\right)$ is
given by $K_2\left(\frac{2\left| \Delta \epsilon
\right|z}{v_f}\right) \simeq \sqrt{\frac{\pi v_f}{ 4 \left| \Delta
\epsilon \right| z}}e^{\frac{-2\left| \Delta \epsilon
\right|z}{v_f}}$. Therefore, we get
\begin{equation}
k_1\simeq\frac{e^2 \mu_{eg}^2  {\left| \Delta \epsilon
\right|}^{3/2}}{96 \sqrt{2}\Omega \hbar^2 \epsilon^2
\pi^{3/2}v_f^{3/2} z^{5/2}} e^{\frac{-4\left| \Delta \epsilon
\right|z}{v_f}}  \label{ppp}.
\end{equation}

   \(\)   \text{\text{  }}\( \)   \end{enumerate}

\subsection{Behavior of rate constant, $k_2$}

We now analyze $k_{2}$. As before, we consider two separate cases.
\(\)   \text{\text{  }}\( \)   \begin{enumerate}
\item
Case I: We consider the case $E_F<\frac{\hbar \Omega}{2}$.
Therefore, $2E_F-\hbar \Omega<0$ and hence $k_2$ is given by
\begin{equation}
\begin{array}{c}
k_2=\frac{e^2 \mu_{eg}^2}{48 \hbar \epsilon^2
\pi^2}\int\limits_{0}^{\infty}dq q^3 e^{-2qz}\frac{\Theta
(qv_{f}-\hbar \Omega)}{\sqrt{q^{2}v_{f}^{2}- \left( \hbar \Omega
\right) ^{2}}}\int\limits_{1}^{\frac{2 E_F+\hbar \Omega}{qv_f}}d\mu
\sqrt{\mu^2-1} \Theta \left[\frac{2 E_F+\hbar
\Omega}{qv_f}-1\right].
\end{array}
\end{equation}
Using the two theta functions in the above expression, $k_2$ can be
written as
\begin{equation}
\begin{array}{c}
k_2=\frac{e^2 \mu_{eg}^2}{48 \hbar \epsilon^2
\pi^2}\int\limits_{\frac{\hbar\Omega}{v_f}}^{\frac{2 E_F+\hbar
\Omega}{v_f}}dq \frac{q^3 e^{-2qz}}{\sqrt{q^{2}v_{f}^{2}- \left(
\hbar \Omega \right) ^{2}}}\int\limits_{1}^{\frac{2 E_F+\hbar
\Omega}{qv_f}}d\mu \sqrt{\mu^2-1}.
\end{array}
\end{equation}
We now make a change of the variable of integration from $q$ to $x$
defined by $q=\frac{\hbar\Omega}{v_f}+x$. Using this transformation,
$k_2$ can be written as
\begin{equation}
\begin{array}{c}
k_2=\frac{e^2 \mu_{eg}^2}{48 \hbar \epsilon^2 \pi^2}e^{ \frac{-2
z\hbar\Omega}{v_f}}\int\limits_{0}^{\frac{2 E_F}{v_f}}dx
\frac{\left(\frac{\hbar\Omega}{v_f}+x\right)^3
e^{-2zx}}{\sqrt{x^{2}v_{f}^{2}+2\hbar\Omega v_f
x}}\int\limits_{1}^{\frac{2 E_F+\hbar \Omega}{xv_f+\hbar
\Omega}}d\mu \sqrt{\mu^2-1}.
\end{array}
\end{equation}
For large values of $z$, in the above integral over $x$, because of
the presence of $e^{-2zx}$ term, only small values of $x$ are
important. Therefore, for small values of $x$, the above expression
can be simplified to get
\begin{equation}
\begin{array}{c}
k_2=\sqrt{\frac{e^4 \mu_{eg}^4 \hbar^3 \Omega^5}{ \pi^4 \epsilon^4
v_f^7}}\frac{e^{ \frac{-2
z\hbar\Omega}{v_f}}}{48\sqrt{2}}\int\limits_{1}^{1+\frac{2
E_F}{\hbar \Omega}}d\mu \sqrt{\mu^2-1}\int\limits_{0}^{\frac{2
E_F}{v_f}}dx \frac{e^{-2zx}}{\sqrt{x}}.
\end{array}
\end{equation}
The integral over $\mu$ can now be performed, and the upper limit in
the integral over $x$ can be extended to $\infty$ to get the
following expression for $k_2$:
\begin{equation}
\begin{array}{c}
k_2=\sqrt{\frac{e^4 \mu_{eg}^4 \hbar^3 \Omega^5}{ \pi^3 \epsilon^4
v_f^7}}\frac{e^{ \frac{-2 z\hbar\Omega}{v_f}}}{192\sqrt{z}}\left(
r\sqrt{r^2-1}-\log \left[ r+\sqrt{r^2-1}\right] \right)
\label{k2long},
\end{array}
\end{equation}
where $r=1+\frac{2 E_F}{\hbar \Omega}$.
\item
Case II: We now consider the case $E_F>\frac{\hbar \Omega}{2}$.
Therefore, $2E_F-\hbar \Omega>0$ and hence $k_2$ is given by
\begin{equation}
\begin{array}{c}
k_2=\frac{e^2 \mu_{eg}^2}{48 \hbar \epsilon^2
\pi^2}\int\limits_{0}^{\infty}dq q^3 e^{-2qz}\frac{\Theta
(qv_{f}-\hbar \Omega)}{\sqrt{q^{2}v_{f}^{2}- \left( \hbar \Omega
\right) ^{2}}}\int\limits_{Max[1,\frac{2 E_F-\hbar
\Omega}{qv_f}]}^{\frac{2 E_F+\hbar \Omega}{qv_f}}d\mu \sqrt{\mu^2-1}
\Theta \left[\frac{2 E_F+\hbar \Omega}{qv_f}-1\right].
\end{array}
\end{equation}
The above equation can be simplified to get
\begin{equation}
\begin{array}{c}
k_2=\frac{e^2 \mu_{eg}^2}{48 \hbar \epsilon^2
\pi^2}\int\limits_{\frac{\hbar\Omega}{v_f}}^{\frac{2 E_F+\hbar
\Omega}{v_f}}dq \frac{q^3 e^{-2qz}}{\sqrt{q^{2}v_{f}^{2}- \left(
\hbar \Omega \right) ^{2}}}\int\limits_{Max[1,\frac{2 E_F-\hbar
\Omega}{qv_f}]}^{\frac{2 E_F+\hbar \Omega}{qv_f}}d\mu
\sqrt{\mu^2-1}.
\end{array}
\end{equation}
We now use the same procedure as was used for evaluating the
integrals in Case I to get
\begin{equation}
\begin{array}{c}
k_2=\sqrt{\frac{e^4 \mu_{eg}^4 \hbar^3 \Omega^5}{ \pi^4 \epsilon^4
v_f^7}}\frac{e^{ \frac{-2
z\hbar\Omega}{v_f}}}{48\sqrt{2}}\int\limits_{Max[1,\frac{2
E_F}{\hbar \Omega}-1]}^{1+\frac{2 E_F}{\hbar \Omega}}d\mu
\sqrt{\mu^2-1}\int\limits_{0}^{\frac{2 E_F}{v_f}}dx
\frac{e^{-2zx}}{\sqrt{x}}.
\end{array}
\end{equation}
On evaluating the integral over $\mu$, we get
\begin{equation}
\begin{array}{c}
k_2=\sqrt{\frac{e^4 \mu_{eg}^4 \hbar^3 \Omega^5}{ \pi^4 \epsilon^4
v_f^7}}\frac{e^{ \frac{-2 z\hbar\Omega}{v_f}}}{96\sqrt{2}}\left(
-s\sqrt{s^2-1}+r\sqrt{r^2-1}+\log \left[ s+\sqrt{s^2-1}\right] -\log
\left[ r+\sqrt{r^2-1}\right] \right) \times\\
\int\limits_{0}^{\frac{2 E_F}{v_f}}dx \frac{e^{-2zx}}{\sqrt{x}},
\end{array}
\end{equation}
where $s=Max[1,\frac{2 E_F}{\hbar \Omega}-1]$ and $r=1+\frac{2
E_F}{\hbar \Omega}$. For $E_F$ close to $\frac{\hbar \Omega}{2}$,
$s=1$. Using this and extending the upper limit of the integral over
$x$ to $\infty$ and then evaluating the integral, we get
\begin{equation}
\begin{array}{c}
k_2=\sqrt{\frac{e^4 \mu_{eg}^4 \hbar^3 \Omega^5}{ \pi^3 \epsilon^4
v_f^7}}\frac{e^{ \frac{-2 z\hbar\Omega}{v_f}}}{192\sqrt{z}}\left(
r\sqrt{r^2-1}-\log \left[ r+\sqrt{r^2-1}\right] \right)
\label{k2long1} .
\end{array}
\end{equation}

   \(\)   \text{\text{  }}\( \)   \end{enumerate}

\end{document}